\documentclass[12pt,preprint]{aastex631}
\usepackage{epsfig}
\usepackage{lineno}
                                                                                
\def\vkm{km s$^{-1}$}

\def\arcs#1{$#1''$}
\def\arcsa#1#2{$#1^{\prime\prime}_{^\textrm{.}}#2$}

\def\solarmass{$M_\odot$}

\def\mrate{$M_\odot$ yr$^{-1}$}

\def\cmc{cm$^{-3}$}

\def\ts#1{\textrm{\scriptsize #1}}

\def\Req{R_{\rm eq}}
\def\Ms{M_\textrm{\scriptsize sd}}

\def\mHa{m_\textrm{\scriptsize H}}

\def\nHt{n_{\textrm{\scriptsize H}_2}}

\def\H2{H$_2$}
\def\N2HP{N$_2$H$^+$}

\def\NH3{NH$_3$}

\def\mH2{m_{\textrm{\scriptsize H}_2}}

\def\putfiga#1#2#3{\epsfig{scale=#1,angle=#2,figure=#3}}
\def\putfig#1#2#3{}
\def\leftblank#1{}

\begin{document}


\title{Angular Momentum Evolution from Core to Disk Scales in the Early Phase of
Star Formation: Constraints from HH 212, HH 211, and B335}

\author{Chin-Fei Lee}
\affiliation{Academia Sinica Institute of Astronomy and Astrophysics,
No.  1, Sec.  4, Roosevelt Road, Taipei 106216, Taiwan, R.O.C.;
cflee@asiaa.sinica.edu.tw}


\begin{abstract}

I investigate the angular momentum evolution from core to disk scales in the
early phase of star formation using three protostellar systems: HH 212, HH
211, and B335.  Observations show that the specific angular momentum follows
a power-law dependence at core scales and transitions to an approximately
constant value at smaller radii, indicating dynamical collapse in the inner
envelope.  I model this behavior using the inside-out collapse solution of
Shu (1977) and its rotating extension, the Terebey-Shu-Cassen model,
including modest magnetic effects through a flattened, magnetized core.  I
find that the observed angular momentum profiles in all three sources
are broadly consistent with an inside-out collapse scenario with
approximate conservation of specific angular momentum in the collapsing
region.  The inferred collapse ages and mass infall rates yield total
masses accreted onto the center that are broadly consistent with the central
masses derived from kinematics, allowing for a fraction of the material to
be ejected by jets and winds.  The predicted midplane densities at typical
radii, including the effect of magnetic flattening, are also consistent with
observational estimates.  The three systems exhibit a similar
pattern of angular momentum evolution
despite large differences in the magnitude of their specific angular momentum.
In particular, the small disk in B335  can be
naturally explained by its lower initial specific angular momentum, although
alternative explanations, such as magnetic braking or different initial
conditions, cannot be excluded. These results suggest that inside-out
collapse, with modest magnetic modification, provides a plausible
first-order description of angular momentum evolution from core to disk
scales in the early phase of star formation.

\end{abstract}

\keywords{stars: formation --- ISM: individual: HH 212, HH 211, B335 ---
ISM: accretion and accretion disk}

\section{Introduction}

The collapse of dense molecular cloud cores into protostars is a fundamental
process in star formation, yet its detailed physical description remains
uncertain.  In the classical framework, inside-out collapse of a singular
isothermal sphere \citep{Shu1977} produces a self-similar infall solution
with a constant mass accretion rate.  When rotation is included, the
Terebey-Shu-Cassen model \cite[hereafter TSC model]{Terebey1984} predicts
that material approximately conserves its specific angular momentum during
infall and forms a centrifugally supported disk at small radii.  While these
models provide a useful baseline, they assume idealized initial conditions
and neglect the effects of magnetic fields and turbulence.

Magnetic fields introduce additional complexity to this picture.  The
initial singular isothermal sphere is generalized to a singular isothermal
toroid \citep{Li1996}.  In the magnetized collapse of such a structure
\citep{Allen2003}, the collapse remains self-similar and closely resembles
the hydrodynamic inside-out solution in its radial structure.  However,
magnetic forces redirect infalling material toward the equatorial plane,
forming a dense pseudodisk \citep{Galli1993}, and introduce modest meridional motions.  In the
presence of rotation, magnetic braking can reduce rotational velocities and
suppress the formation of large Keplerian disks \citep{Allen2003rot}.  More
generally, recent theoretical studies have shown that the efficiency of disk
formation depends sensitively on the transport of angular momentum by
magnetic fields and on processes such as non-ideal MHD effects,
magnetic-field misalignment, and turbulence, which can weaken magnetic
braking and facilitate disk growth \citep{Zhao2020,Tsukamoto2023}.  At the
same time, recent high-resolution observations have revealed a wide
diversity of disk sizes and kinematic structures in the youngest protostars
\citep{Segura-Cox2018,Tobin2020,Maury2019,Hsieh2024},
thereby providing increasingly stringent tests of these theoretical predictions
\citep{Zhao2020,Tsukamoto2023}.
Despite these effects, the overall density and radial velocity structures
are not drastically altered compared to the non-magnetic case, and the mass
infall rate is not strongly enhanced in the rotating magnetized solutions. 
Thus, magnetized collapse provides a mechanism for redistributing angular momentum
without fundamentally changing the global collapse
kinematics. 

Observational constraints on angular momentum in dense cores and
protostellar envelopes have improved significantly over the past decades. 
Early studies by \citet{Goodman1993} showed that the specific angular
momentum in dense cores scales with radius approximately as $j(r) \propto
r^{1.6\pm0.2}$.  More recently, \citet{Chen2019} found that the observed
distribution of specific angular momentum, based on a larger sample of dense
cores, can be reasonably described by a power-law relation with an index of
$\sim$1.5.  Both are consistent with expectations from turbulent motions in
molecular clouds, as implied by the linewidth-size relation
\cite[e.g.,][]{Larson1981,Solomon1987} and supported by numerical models
\cite[e.g.,][]{Burkert2000}.  Using an even larger sample of 329 dense
cores, although spanning a smaller range of spatial scales,
\citet{Pandhi2023} derived a best-fit relation, $j(r)\propto r^{1.8\pm0.1}$,
suggesting that the observed velocity gradients are likely produced by a
combination of solid-body rotation and turbulent motions.  
Interferometric observations by \citet{Ohashi1997} of Class I objects
indicated a flattening of the angular momentum profile at radii of order
$\sim1000$ au, possibly marking the transition to the collapsing region. 
Since these sources are more evolved than the Class 0 objects studied
here, a larger transition radius is expected.
More recent measurements by \citet{Pineda2019}
found $j(r)\propto r^{1.8}$ over scales of $10^3-10^4$ au, primarily tracing
the core and outer envelope.  Complementary results from \citet{Gaudel2020}
show a similar power-law dependence in the outer envelope, $j(r) \propto
r^{1.6\pm0.2}$, transitioning to an approximately constant value inside
$\sim$ 1600 au, suggesting entry into the collapsing regime.  Overall, these
observational studies indicate that the specific angular momentum in dense
cores approximately follows a power-law relation with an index of $\sim
1.5-1.8$.

Individual protostellar systems provide further insight into the angular
momentum evolution from core to envelope scales.  The isolated protostar
B335 shows extremely weak rotation and only a very small or unresolved disk
\citep{Kurono2013,Yen2015}, although it remains unclear whether this
reflects efficient angular momentum removal by magnetic braking or simply
the young evolutionary stage of the source.  The protostellar system HH 211
exhibits a more organized rotational structure, with a specific angular
momentum profile from the core to the infalling envelope that can be naturally
explained by inside-out collapse \citep{Lee2019HH211}.  Rather than
representing fundamentally different cases, these systems may reflect
different initial conditions or evolutionary stages, highlighting the need
for a systematic examination of angular momentum evolution across multiple
sources.

Taken together, these observational results suggest that while the angular
momentum distribution at core scales is broadly consistent with a turbulent
origin, its evolution within the collapsing region remains uncertain.  In
particular, it is not yet clear whether the classical inside-out collapse
framework, with or without magnetic effects, can fully reproduce the
observed radial profiles of specific angular momentum and the diversity of
disk properties.

In this paper, I investigate the angular momentum structure in the early
phase of star formation using existing observations from the literature.  I
first outline the inside-out collapse model and then re-examine previous
results for HH~212, HH~211, and B335.  By comparing these systems, I
investigate whether inside-out collapse can reproduce the observed radial
profiles of specific angular momentum from core to disk scales, as well as
the central masses and midplane densities derived from the observations.


\section{Inside-out Collapse Model}

I start with the inside-out collapse model of \cite{Shu1977}. In this model, the
core is initially a singular isothermal sphere in hydrostatic equilibrium
between thermal pressure and gravity, with a density profile set by the
isothermal sound speed $a$:
\begin{equation}
\rho(r) = \frac{a^2}{2\pi G r^2} 
\end{equation}
corresponding to a molecular hydrogen number density
\begin{equation}
\nHt(r) = \frac{\rho(r)}{2.8 \mHa}
\end{equation} 
An expansion wave propagates outward at the sound speed, initiating collapse
within its radius with an infall
rate
\begin{equation}
\dot{M} = 0.975 \frac{a^3}{G}
\end{equation} 
The collapse radius ($r_c$) marks the position of the expansion wave, and
the collapse age ($t_c$) corresponds to the time since its launch.  The
self-similar solution provides the infall velocity and density
\citep{Shu1977}.  In the inner region, pressure gradients become negligible
compared to gravity, and the flow approaches ballistic
free fall, with the density scaling approximately as $r^{-1.5}$.

\citet[hereafter TSC]{Terebey1984} extended this model to include slow,
solid-body rotation.  At radii much larger than the centrifugal radius, the
collapse follows the non-rotating solution, with angular momentum advected
inward.  As the collapse proceeds inward, the flow approaches ballistic
motion, approximately conserving energy and specific angular momentum along
streamlines.  More importantly, in this ballistic infall regime, material
originates from a narrow range of initial radii and thus carries
approximately the same specific angular momentum (TSC).  In this regime, the
flow can be approximated by the \citet{Ulrich1976} solution. 
In the equatorial plane, the infall velocity and density in the TSC model
can be approximated as 
\begin{equation} 
v^\ts{T}_{r,\textrm{\scriptsize
mid}}(r) \approx v^\ts{S}_r (r)(1-\frac{R_c}{2r})^{1/2} \label{eq:vrTSC}
\end{equation} 
\begin{equation} 
n_{\textrm{\scriptsize
H}_2,\textrm{\scriptsize mid}}^\ts{T}(r) \approx \nHt^\ts{S}(r)
(1-\frac{R_c}{2r})^{-1/2}(1-\frac{R_c}{r})^{-1} \label{eq:denTSC}
\label{eq:TSCU}
\end{equation}
where $v^\ts{S}_r(r)$ and $\nHt^\ts{S}(r)$ are given by the
Shu solution. 
The centrifugal radius is defined as
$R_c=\frac{j^2}{G\Ms}$,
where $j$ is the specific angular momentum of the infalling material and
$\Ms$ is the central mass (protostar plus disk) that governs the
gravitational potential. It is the radius at which gravity and centrifugal
acceleration balance for material with a given specific angular momentum.
The radius of the centrifugal barrier is defined as
$R_b=\frac{j^2}{2G\Ms}$,
where the radial velocity of the infalling material becomes zero for a given specific angular momentum.
At this
location, the gravitational potential energy is completely converted
into rotational kinetic energy in the ballistic solution. Therefore, the
radius of the centrifugal barrier is equal to one-half of the centrifugal radius.
As shown in Eq. \ref{eq:TSCU}, rotation reduces the infall velocity and enhances the density
through slower infall and streamline convergence toward the midplane.  The
divergence at the centrifugal radius is an artifact of the ballistic
approximation, which neglects pressure gradients, turbulence, and magnetic
effects.

While the TSC solution formally extends to the centrifugal radius, numerical
simulations \cite[e.g.,][]{Das2026} show that the infall velocity decreases
to zero at the centrifugal barrier, where a Keplerian disk forms.  Assuming
that energy and angular momentum are approximately conserved down to this
barrier, modeling the kinematics of the infalling envelope provides
constraints on the central mass and specific angular momentum 
\cite[e.g.,][]{Sakai2014,Lee2017HH212}.  Additional constraints on the central mass
can be obtained from the disk kinematics.

The initial core is likely magnetized, with poloidal magnetic fields
producing a flattened structure \citep{Girart2006,Myers2020,Yen2021,Ward2023,Huang2024}.  \citet{Li1996} showed that a magnetized
core in quasi-equilibrium can be described as a singular isothermal toroid
with a density
\begin{equation}
\rho(r) = \frac{a^2}{2\pi G r^2} R(\theta)
\end{equation}
where $R(\theta)$ characterizes the angular dependence and
$\theta=\pi/2$ corresponds to the equatorial plane.
The midplane density is therefore enhanced by a factor of 
$\Req\equiv R (\pi/2)$ relative to the non-magnetized case.
 The resulting toroid has an elongation $\chi$ (defined as the
width-to-height ratio) that depends on the parameter $H_0$, which quantifies
the degree of magnetic support, and thus the corresponding mass-to-flux
ratio $\lambda$.  The observed cores discussed later in this paper exhibit
elongations of $\chi \sim 2$--3, corresponding to $H_0 \sim 0.375$--0.75 and
$\lambda \sim 2$--3 according to Table 1 of \citet{Li1996}.  For such
moderately magnetized cores, the initial density structure is modestly
flattened, resulting in an initial midplane density enhanced by a factor of
$\Req \sim 2$--4 \cite[interpolating from Figure 1b in][for the required $H_0$ value]{Li1996} relative
to the non-magnetized configuration.


The subsequent collapse remains approximately self-similar
\citep{Allen2003,Allen2003rot} and is broadly analogous to that of the
TSC solution, although it proceeds from a magnetized toroidal initial
condition rather than a singular isothermal sphere.  During collapse,
magnetic fields redirect infalling material toward the equatorial plane,
transforming the flattened toroid into a dynamically infalling pseudodisk. 
This process primarily redistributes the flow and increases the degree of
flattening \citep{Allen2003rot}.  The midplane density in the magnetized
collapse solution is therefore expected to remain enhanced by approximately
a factor of $\Req$ relative to that in the TSC solution. Despite these
structural changes, the mass infall rate remains roughly the same as that in
the TSC model \cite[see their Figure 2 in][]{Allen2003rot}.

\section{HH 212}

\subsection{Observational Results}



In this paper, I adopt a distance of 400 pc for HH 212. 
\citet{Wiseman2001} detected a flattened, rotating ammonia core around the
central source with an  elongation of $\sim$ 2, extending $\sim$ 8000 au
(\arcs{20}) to the southeast and northwest.  As shown in their Figure 2, the
radial velocity profile along the equatorial plane has been measured out to
a radius of 4560 au (\arcsa{11}{4}), with a central velocity of $\sim$ 1.6
\vkm{}.  I adopt this value as the systemic velocity of the core to derive
its rotation and specific angular momentum.  This is consistent with the
systemic velocity of the HCO$^+$ envelope, which is $\sim$ 1.7$\pm$0.1
\vkm{} \citep{Lee2014HH212,Lee2017HH212}.

\begin{figure} [!hbp]
\centering
\putfiga{0.65}{270}{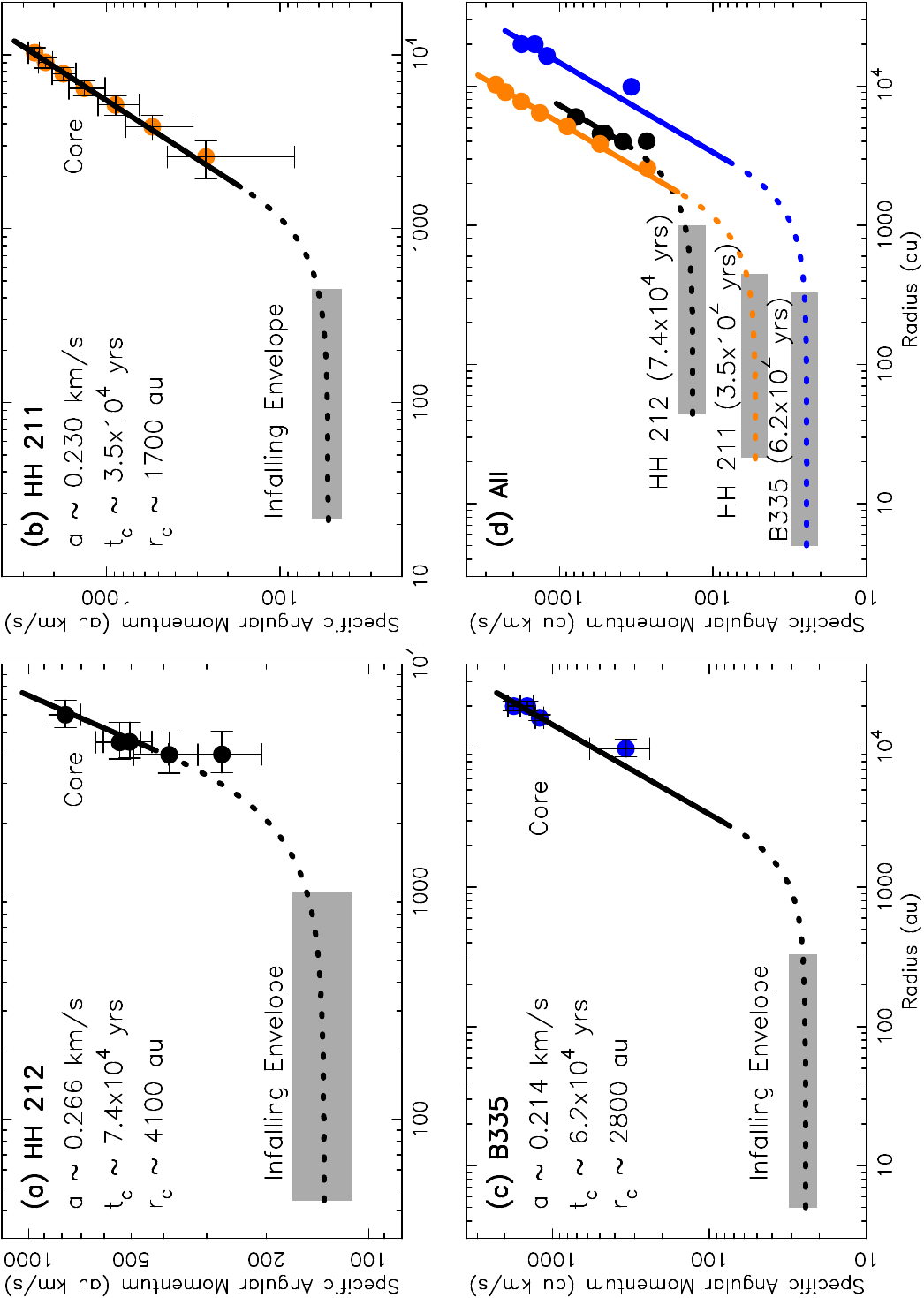} 
\figcaption[]
{
Distributions of specific angular momentum from core to infalling envelope
in (a) HH 212, (b) HH 211, and (c) B335, compared with the inside-out
collapse model (solid and dotted lines). 
Error bars in panels (a)-(c) represent the uncertainties in the measurements.
Panel (d) shows a comparison among
the three sources, with the error bars omitted for clarity.
In all panels, dots represent the observed distributions
in the core, while shaded regions indicate those in the infalling envelope. 
The solid lines trace the outer core that remains static prior to collapse, following a
power-law distribution.  The dotted lines show the model
predictions for the inside-out collapse scenario,
illustrating the transition from the collapsing inner core to the infalling
envelope.
\label{fig:lr}}
\end{figure}

Figure \ref{fig:lr}a shows the distribution of specific angular momentum
from the core to the infalling envelope for HH 212.  Given the angular
resolution of the ammonia observations ($\sim$ 3360$\times$3520 au), I
derive the specific angular momentum of the core at radii of 4000 and 4560
au, shown as black circles in the figure.  Adopting a mean
velocity gradient of $\sim$ 4.5 \vkm{} pc$^{-1}$ across the core diameter of
12,000 au \citep{Wiseman2001}, I further estimate a specific angular
momentum of $\sim$ 780 au \vkm{} near the core edge at 6000 au.  
The uncertainties are assumed to
be constant for all data points, with a positional uncertainty of
one-quarter of the beam size ($\sim$ 830 au) and a specific angular momentum
uncertainty of $\sim$ 83 au \vkm{}, estimated as one-quarter of the beam size
multiplied by one-third of the velocity resolution.
The figure
also shows the specific angular momentum distribution of the infalling
envelope traced in HCO$^+$ within $\sim$ 1000 au (shaded region)
\citep{Lee2014HH212,Lee2017HH212}, derived assuming ballistic motion with
conservation of total energy and constant angular momentum.  The specific
angular momentum is $\sim140\pm30$ au \vkm{}, with a central mass of $\sim$
0.25$\pm$0.05 \solarmass{}, implying a centrifugal barrier at $\sim$ 44 au
and a centrifugal radius of $\sim$ 88 au. A rotating disk is detected
at the centrifugal barrier \citep{Lee2017HH212D,Lee2017HH212}.
As shown in the figure, the
specific angular momentum decreases with decreasing radius in the core and
transitions to an approximately constant value in the infalling envelope, as
expected in the inside-out collapse model (TSC).


\subsection{Model Parameters and Comparison}



In the inner envelope, the infall rate has been estimated to have a mean
value of $\sim 4.3\times10^{-6}$ \mrate{} \citep{Lee2014HH212,Lopez2026}. 
In the Shu model, this corresponds to a temperature of $\sim$ 20 K.  I
therefore adopt a temperature of 20 K, which gives an isothermal sound speed
$a = 0.266$ \vkm{}. 
Although the temperature varies
throughout the core and envelope, the Shu/TSC collapse solution assumes an
isothermal equation of state and is therefore characterized by a single
sound speed. The adopted temperature should thus be regarded as an
effective temperature used to characterize the overall collapse rather than
the local temperature at every location.
The corresponding pre-collapse density at a radius of
4000 au is $\nHt \sim 10^5$ \cmc{}, consistent with the mean density
estimated in the ammonia core \citep{Wiseman2001}.


As reviewed in the Introduction, the specific angular momentum in dense
cores approximately follows a power-law relation with an index ranging from
$\sim$1.5 to $\sim$1.8.
For HH~212, I adopt a power-law
index of 1.55, as in HH~211, which provides a reasonable fit to the observed
core profile,
$j \sim 46 (r/1000\; \textrm{au})^{1.55}$ au \vkm{} (solid line). The fit
was performed using the nonlinear least-squares Marquardt--Levenberg
algorithm implemented in {\tt gnuplot}, with the observed data points used
directly without additional binning. The uncertainties described above were
adopted in the fitting, but since they are identical for all data points,
they do not affect the best-fit parameters.

The specific angular momentum in the original core at smaller radii is
not directly constrained by observations and is therefore assumed to follow the same
profile as that observed in the outer core.  Assuming that the angular
momentum is carried inward with the collapsing material, the specific
angular momentum in the collapsing envelope is approximately equal to that
of the original core at one-half of the collapse radius \citep{Terebey1984},
according to the inside-out collapse solution of \citet{Shu1977}.  Using the
observed specific angular momentum in the collapsing envelope, the collapse
radius is estimated to be $\sim$4100 au.  With the adopted
isothermal sound speed, this corresponds to a collapse age of
$\sim7.4\times10^4$ yr.  The dotted line in the figure illustrates the
collapse profile of the specific angular momentum, and the inferred collapse
radius is given by its intersection with the solid line. For an infall rate
of $\sim$ 4.3$\times10^{-6}$ \mrate, the total mass accreted onto the center
is $\sim$ 0.32 \solarmass{}.  The central mass previously derived from
kinematics corresponds to $\sim$ 80\% of this value.  This is consistent
with theoretical expectations, given that 10$-$30\% of the accreted mass may be ejected
away in a jet and a wind \citep{Shu2000,Pudritz2007,Lee2020}.  Notice
that, if a power-law index of 1.8 is adopted, then the specific angular
momentum of the core becomes $j \sim 31 (r/1000\;\textrm{au})^{1.8}$ au
\vkm{}.  The corresponding collapse radius and collapse age would be about
12\% larger.  The inferred accreted mass would also increase by the same
amount, becoming $\sim$ 0.36 \solarmass{}.  In that case, approximately 30\%
of the accreted mass would need to be ejected.

From the TSC model, the midplane density at a representative radius of 400
au is predicted to be $\sim 1.7\times10^6$ \cmc{}.  The core is likely
initially a flattened, magnetized toroid, as it is magnetized
\citep{Yen2021} and exhibits an  elongation of $\sim$ 2
\citep{Wiseman2001}.  According to \citet{Li1996}, this elongation
corresponds to $H_0 \sim 0.375$ and a mass-to-flux ratio of $\lambda \sim
3.2$, yielding $\Req \sim 2.3$.  Applying this $\Req$ value to the TSC
model yields a midplane density of $\sim 3.9 \times 10^6$ cm$^{-3}$ at 400
au, consistent with the observational estimate of $\sim 5.0 \times 10^6$
cm$^{-3}$ to within about 25\%.  The observational estimate was derived
by fitting the structure, spectrum, and kinematics of the HCO$^+$ emission
using a 3D radiative transfer model of the collapsing envelope
\citep{Lee2014HH212}.

\section{HH 211}


\subsection{Observational results}

Figure \ref{fig:lr}b shows the distribution of specific angular momentum
from the core to the infalling envelope for HH 211.  I adopt a distance of
321 pc for this source.  The dense core, traced in ammonia, has a size of
$\sim$ 20000 au $\times$ 6000 au and an elongation of $\sim$ 3, with its
specific angular momentum measured over radii of $\sim$ 2600 to 10000 au
(see the orange circles) \citep{Tanner2011}.  The infalling envelope has
been detected in C$^{18}$O within $\sim$ 400 au of the central source by
\citet{Lee2019HH211}.  Modeling the envelope with a ballistic model, they
derived a central mass of $\sim$ 0.08$\pm0.02$ \solarmass{} and a specific
angular momentum of 55$\pm15$ au \vkm{} (see the shaded region), implying a
centrifugal barrier at 21 au and a centrifugal radius of 42 au. 
A compact rotating disk is observed
at the centrifugal barrier \citep{Lee2018HH211,Lee2023HH211}.
As can be
seen, the distribution of specific angular momentum from the core to the
infalling envelope follows a profile similar to that seen in HH 212.

\subsection{Model parameters and comparison}

The ammonia core has a mean temperature of $\sim$ 15 K \citep{Tanner2011}. 
Using this temperature, the isothermal sound speed is $a =0.233$ \vkm{}. 
This yields a density of $\sim$ $10^5$ \cmc{} at 3500 au, broadly consistent
with the observed mean core density of $(1-2)\times10^5$ \cmc{}
\citep{Yen2023}.  This also yields an infall rate of $\sim$
2.8$\times10^{-6}$ \mrate{}.

The observed specific angular momentum in the core has been described by $j
\sim 72 (r/1000\; \textrm{au})^{1.55}$ au \vkm{} (see the solid line)
\citep{Lee2019HH211}, significantly larger than that in HH 212.  As for
HH 212, using the observed specific angular momentum in the collapsing
envelope, the collapse radius is estimated to be $\sim$1700 au (see the
dotted line).  With the adopted sound speed, the collapse age is thus $\sim$
3.5$\times10^4$ yrs.  The dotted line in the figure illustrates the collapse
profile of the specific angular momentum. A highly pinched poloidal field
in the innermost envelope \citep{Lee2019HH211} and an hourglass field
morphology in the outer envelope \citep{Choi2025} are detected within this
radius, supporting collapse from a magnetized core.  The total mass accreted
onto the center is thus $\sim$ 0.10 \solarmass{}, of which the central mass
derived from kinematics is $\sim$ 80\%.  This is expected if 10$-$30\% of
the mass is removed by a jet and a wind \citep{Shu2000,Pudritz2007,Lee2020}.

From the TSC model, the midplane density at a representative radius of 100
au is $\sim 2.0\times10^7$ \cmc{}.  The core is likely initially a
flattened, magnetized toroid, as it is magnetized \citep{Yen2023} and
exhibits an elongation of $\sim$ 3 \citep{Tanner2011}.  According to
\citet{Li1996}, this elongation corresponds to $H_0 \sim 0.75$ and a
mass-to-flux ratio of $\lambda \sim 2.1$, yielding $\Req \sim 4.2$. 
Applying this $\Req$ value to the TSC model yields an expected
midplane density of $\sim 8.4 \times 10^7$ cm$^{-3}$ at 100 au. 
 
Observationally, the midplane density can also be estimated from the
observed column density to compare with the model prediction. In
HH~211, the envelope is nearly edge-on, so the column density at a
projected radius $r_0$ can be obtained by integrating the density along
the line of sight out to the collapse radius,
\begin{equation}
N(r_0)=2\int_{r_0}^{r_c}
n_0\left(\frac{r}{r_0}\right)^{-1.5}
\frac{r}{\sqrt{r^2-r_0^2}}\,dr,
\end{equation}
where $n_0$ is the midplane density at $r_0$, assuming a density profile
$n\propto r^{-1.5}$ as predicted by the collapse solution. For
$r_0=100$ au, the integral yields
$N(r_0)\approx4.2\,r_0\,n_0$, implying an effective line-of-sight path
length of $4.2\,r_0\approx420$ au. At this projected radius, the
observed column density was estimated to be $\sim2.43$ g cm$^{-2}$ from
the dust continuum emission, assuming optically thin emission and a dust
temperature of $\sim40$ K \citep{Lee2019HH211}. This column density thus
implies a midplane density of
$n_0\sim8.3\times10^7~{\rm cm}^{-3}$ at 100 au, consistent with the model
prediction. In contrast, \citet{Lee2019HH211} estimated the midplane
density by dividing the observed column density by a path length of
100 au instead of the effective path length of $\sim420$ au. This
approximation underestimates the path length by a factor of
$\sim4.2$ and consequently overestimates the midplane density by the
same factor.

\section{B335}

\subsection{Observational Results}

Figure \ref{fig:lr}c shows the distribution of specific angular momentum
from the core to the infalling envelope for B335.  I adopt a distance of
$\sim$ 165 pc for B335, corresponding to the Gaia distance to HD 184982
associated with the cloud \citep{Watson2020}.  \citet{Saito1999} observed a
flattened core in H$^{13}$CO$^+$ and C$^{18}$O with an {elongation of $\sim$ 2},
extending to 25000 au ($\sim$ \arcs{150}) along the north-south direction. 
The blue circles show the specific angular momentum of the core measured in
C$^{18}$O at 20000 au \citep{Saito1999} and at 9900 au \citep{Yen2010}, and
in HCO$^{+}$ at 16500 and 20000 au \citep{Kurono2013}, adjusted to the
adopted distance.  

\begin{figure} [!hbp] \centering \putfiga{0.6}{270}{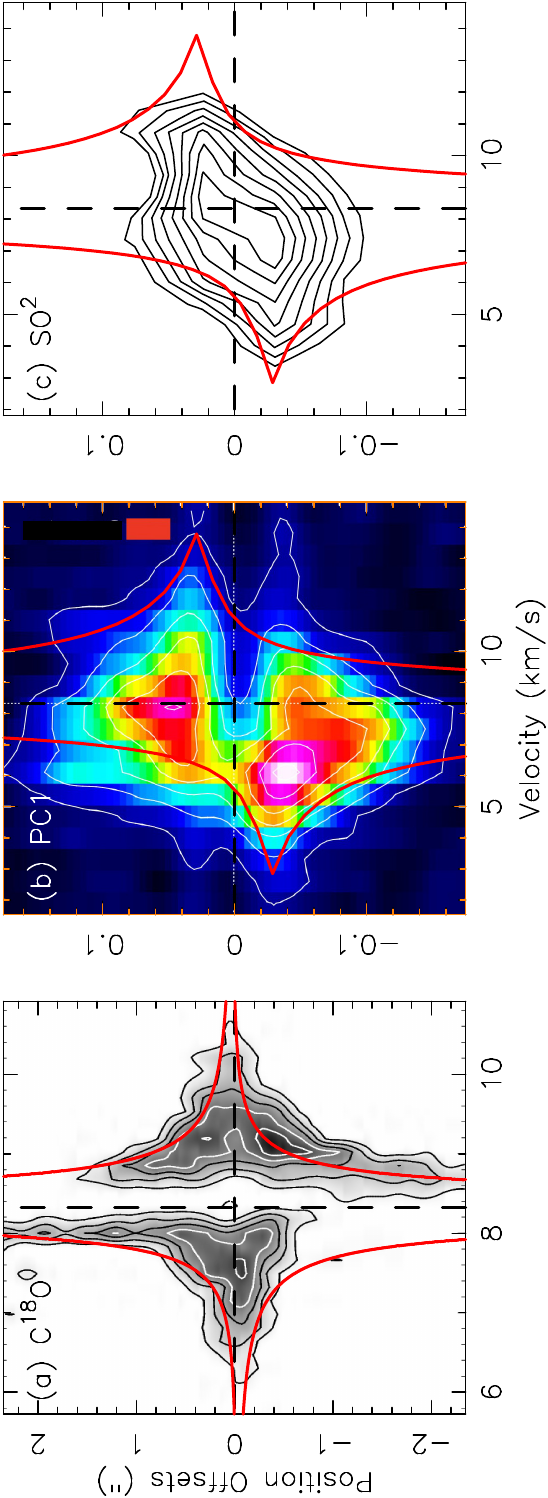} 
\figcaption[] { Position-velocity (PV) diagrams of the infalling envelope in
B335 at different spatial scales along the major axis in (a) C$^{18}$O
\citep{Yen2015}, (b) the first principal component (PC1) from complex
organic molecules \citep{Okoda2022}, and (c) SO$_2$ \citep{Bjerkeli2019}.  
The red curves are calculated using the simple ballistic model of
\citet{Sakai2014}, which assumes conservation of energy and specific angular
momentum in the infalling material, adopting $\Ms \sim 0.08$ \solarmass{} and
$j \sim 26$ au \vkm{}.
The vertical dashed lines mark the systemic velocity. The horizontal dashed lines mark the
position of the central source.
\label{fig:pvB335}}
\end{figure}

The inner collapsing region is detected within $\sim$ 350 au of the central
source.  I re-examine the position-velocity (PV) diagrams along the major
axis previously reported in C$^{18}$O \cite[Figure 5 in][]{Yen2015}, the
first principal component (PC1) from complex organic molecules \cite[Figure
6 in][]{Okoda2022}, and SO$_2$ \cite[Figure 8 in][]{Bjerkeli2019}, as shown
in Figure \ref{fig:pvB335}. 
The PV diagrams can be modeled simultaneously with the simple ballistic
model introduced by \citet{Sakai2014}, in which the rotation velocity is
given by $v_\phi=l/r$ and the infall velocity by
$v_r=-\sqrt{2G\Ms/r-v_\phi^2}$, where $l$ is the specific angular
momentum of the infalling material and $\Ms$ is the central mass (protostar plus disk).  
By visually comparing the model predictions with the observed PV diagrams and
requiring that all three diagrams be reproduced simultaneously,
I find a central mass of
$\sim0.08\pm0.02$ \solarmass{} and a specific angular momentum of
$\sim26\pm8$ au \vkm{} (shaded region).
The quoted uncertainties were estimated by varying the central mass and
specific angular momentum to determine the range of values that still
provided a reasonable visual match to all three observed PV diagrams
simultaneously.
This mass is consistent with previous estimates
\citep{Yen2015,Bjerkeli2019,Okoda2022}.  The derived specific angular
momentum lies between the values reported by \citet[$\sim$ 10 au
km/s]{Yen2015} and \citet[$\sim$ 40 au \vkm{}]{Bjerkeli2019}.  The resulting
centrifugal barrier is $\sim$ 5 au, consistent with that derived by
\citet{Imai2019}.  A dusty disk is observed at the centrifugal
barrier \citep{Bjerkeli2019}.  This therefore suggests that SO$_2$, which
exhibits a roughly linear PV structure, traces a shocked ring near the disk
edge.

\subsection{Model Parameters and Comparison}

The density profile of the core has been modeled using the Shu inside-out
collapse solution based on H$_2$CO emission with a temperature of 13 K
\citep{Zhou1990}, C$^{18}$O emission with 12 K \citep{Saito1999}, and
H$^{13}$CO$^+$ emission with 15 K \citep{Kurono2013}.  Thus I adopt a
temperature of 13 K, as in \citet{Evans2015}, which gives an isothermal
sound speed of 0.214 \vkm{} and an infall rate of $\sim$ 2.3$\times10^{-6}$
\mrate{}. As for HH~212, a single effective temperature is adopted to
characterize the overall collapse.

Assuming the same power-law index as for HH 211, the specific angular
momentum in the core can be obtained with $j \sim 15.4 (r/1000 \;
\textrm{au})^{1.55}$ (see the solid line), significantly lower than that in
both HH 212 and HH 211. 
Using the observed specific angular momentum in the collapsing envelope, the collapse
radius is therefore estimated to be $\sim$2800 au.  With the adopted
isothermal sound speed, this corresponds to a collapse age of
$\sim6.2\times10^4$ yr.  The dotted line in the figure illustrates the
collapse profile of the specific angular momentum, and the inferred collapse
radius is given by its intersection with the solid line.
 ALMA polarization observations reveal a pinched magnetic field morphology
on $\sim$ 1000 au scales \citep{Maury2018}, supporting collapse from a
magnetized core.  The total mass accreted to the center is $\sim$ 0.14
\solarmass{}, of which the central mass derived from kinematics is $\sim$
57\%.  This implies that $\sim$ 40\% of the mass may be removed by a jet and
a wind, somewhat higher but still comparable to typical predicted values of
10-30\% \citep{Shu2000,Pudritz2007}.  Notice that, if a power-law index
of 1.8 is adopted, then the specific angular momentum of the core becomes $j
\sim 7.3 (r/1000\;\textrm{au})^{1.8}$ au \vkm{}.  The corresponding collapse
radius and collapse age would be about 43\% larger.  The inferred accreted
mass would also increase by the same amount, becoming $\sim$ 0.2
\solarmass{}.  In that case, approximately 60\% of the accreted mass would
need to be ejected.  This would be significantly higher than that predicted. 
Since the specific angular momentum in this source is relatively small,
turbulent motions are likely to play a more important role.  Therefore, the
power-law index is likely to be closer to 1.5.


From the TSC model, the midplane density at a representative radius of 100
au is $\sim 7.6\times10^6$ \cmc{}.  The core is likely initially a
flattened, magnetized toroid, as it is magnetized \citep{Yen2023} and
exhibits an  elongation of $\sim$2 \citep{Saito1999}. 
According to \citet{Li1996}, this elongation corresponds to
$H_0 \sim 0.375$ and a mass-to-flux ratio of
$\lambda \sim 3.2$, yielding $\Req \sim 2.3$.
Applying this $\Req$ value
to the TSC model yields a
midplane density of $\sim 1.7 \times 10^7$ cm$^{-3}$ at 100 au, consistent
with the observational estimate of $\sim 1.0 \times 10^7$ cm$^{-3}$
\citep{Yen2015} to within a factor of 2. 
The observational estimate was
derived by fitting the observed position-velocity diagrams of the
C$^{18}$O emission from the collapsing envelope with a 3D radiative
transfer model.
Considering uncertainties in
distance, specific angular momentum, and central mass, the agreement is
therefore satisfactory.

\section{Discussion}

\subsection{A Common Angular Momentum Profile from Core to Envelope}

The three protostellar systems analyzed here -- HH~212, HH~211, and
B335 -- exhibit a similar radial behavior of specific angular momentum from
core to envelope scales, despite large differences in magnitude (see Figure
\ref{fig:lr}d).  In all cases, the specific angular momentum follows a
power-law dependence in the core and transitions to an approximately
constant value in the inner envelope.  This behavior is consistent with
early studies of dense cores (e.g., \citealt{Goodman1993}) and
interferometric observations (e.g., \citealt{Ohashi1997}), and is further
supported by recent measurements (e.g., \citealt{Pineda2019, Gaudel2020}),
which report $j(r) \propto r^{1.5-1.8}$ at core scales and flattening toward
smaller radii.

The present results extend these findings by linking the outer-core power-law
profile to the inner collapsing region within a unified framework.  The
observed transition can be naturally explained by inside-out collapse, in which
material in the inner region originates from a limited range of initial
radii and thus retains an approximately constant specific angular momentum. 
The consistency among the three sources suggests that this interpretation
provides a plausible description of angular momentum evolution during
collapse, although a larger sample will be required to establish its
generality.


\subsection{Magnetic Field Effects: Pseudodisks and Angular Momentum Evolution}

The observed elongations of these cores are interpreted as resulting
from the effect of a large-scale poloidal magnetic field, as modeled by
\citet{Li1996}.  However, this interpretation is not unique, and further
work is needed to exclude other possible origins of the elongation, such as
filaments, streamers, or intrinsically elongated initial core structures.

The infalling envelopes in these systems are best interpreted as pseudodisks
rather than quasi-spherical infall.  Observationally, these structures
exhibit pronounced flattening and show kinematics dominated by infall rather
than rotational support.  In addition, polarization observations reveal
pinched, hourglass-shaped magnetic field morphologies on $\sim 10^3$ au
scales in some sources, consistent with material being channeled toward the
equatorial plane by magnetic fields.  These properties are naturally
explained in magnetized collapse models, in which magnetic fields produce a
flattened pseudodisk structure even in the absence of rotational support
\citep{Galli1993, Allen2003}.

Such pinched field morphologies arise from the dragging of initially
poloidal field lines by the infalling material \citep{Galli1993,Allen2003}. 
Observational evidence for this has been reported in several protostellar
systems, most clearly in HH~211, where polarization observations reveal a
well-defined hourglass morphology \citep{Lee2019HH211, Choi2025}.  In B335,
polarization observations also indicate a pinched field configuration on
similar scales \citep{Maury2018}, while in HH~212 the magnetic field
structure is less well constrained but remains broadly consistent with a
magnetized collapse scenario \citep{Yen2023}.

While these results show that magnetic fields play an important role in
shaping the geometry of the collapsing flow, their impact on angular
momentum transport appears to be limited at this stage.  In all three
sources, the specific angular momentum remains approximately constant in the
inner envelope, indicating that angular momentum is largely conserved during
infall.  This behavior contrasts with predictions from numerical simulations
of magnetized collapse, in which magnetic braking can efficiently remove
angular momentum and suppress disk formation \citep{Mellon2008,
Hennebelle2008, Hennebelle2009, Machida2011, Li2014}.
This conclusion applies only to the three
sources studied here, and selection effects may exist because all three
systems exhibit measurable rotational structures.

The reduced efficiency of magnetic braking may reflect the role of non-ideal
MHD effects, such as ambipolar diffusion, Ohmic dissipation, and the Hall
effect, which can decouple neutral gas from magnetic fields in dense regions
\citep{Li2014, Zhao2016, Wurster2018}.  Observational indications of such
effects have been reported in protostellar systems, including recent
promising evidence for ambipolar diffusion in HH~212 \citep{Lopez2026}.

On the other hand, significant angular momentum loss has been observed in
some more evolved protostellar systems, such as HH~111, where the specific
angular momentum decreases toward smaller radii \citep{Lee2016HH111, Lin2025}. 
This suggests that magnetic braking can become effective under certain
conditions, possibly at later evolutionary stages or in systems with
stronger magnetic coupling.  Together, these results indicate that while
magnetic fields shape the collapse and influence angular momentum transport,
their efficiency varies among sources and may evolve with time.

\subsection{Disk Size Diversity and Early Disk Formation}

The three sources exhibit a wide range of disk sizes, with B335
\citep{Bjerkeli2019} showing a particularly small disk compared to HH~212
\citep{Lee2017HH212D} and HH~211 \citep{Lee2023HH211}.  This diversity can be
naturally understood in terms of differences in the specific angular
momentum and central mass.  In the context of the inside-out collapse model
\citep{Shu1977} and its rotating extension \citep{Terebey1984}, the
characteristic disk size is linked to the centrifugal radius, which scales
as $R_c \propto j^2/\Ms$, while the observed disk boundary is more closely
associated with the centrifugal barrier, where infalling material
transitions to rotational support.  Thus, for a given central mass, a lower
specific angular momentum leads to a smaller centrifugal radius and barrier. 
The significantly lower angular momentum observed in B335 therefore provides
a natural explanation for its small disk size, without invoking enhanced
angular momentum removal \citep{Yen2015, Bjerkeli2019, Okoda2022}.
However,
alternative explanations cannot yet be excluded. In particular, stronger
magnetic coupling could enhance magnetic braking and further suppress disk
growth, while uncertainties in the central mass, source geometry, or
evolutionary stage may also contribute to the inferred disk size.

This interpretation is consistent with observational studies of
protostellar samples, which show substantial variations in specific angular
momentum and inferred centrifugal radii \citep{Goodman1993, Gaudel2020}. 
However, the large uncertainties in central mass estimates and the wide
range of possible centrifugal radii make it difficult to draw firm
conclusions about disk sizes on a source-by-source basis.  In particular,
some studies suggest that the specific angular momentum remains
approximately constant down to radii of tens of au \citep{Gaudel2020},
implying that disk sizes are likely smaller than or comparable to these
scales in the early phase of star formation.
Moreover, recent studies
suggest that the efficiency of magnetic braking depends on the degree of
coupling between the gas and magnetic field, which is itself sensitive to
the local ionization fraction and cosmic-ray ionization rate
\citep{Cabedo2023,Pineda2024}. Consequently, local variations in these
quantities may also influence disk formation and the resulting disk size.

The results presented here suggest that much of the diversity of disk sizes in young
protostars is primarily set by initial conditions, particularly the
distribution of angular momentum in the parent core, rather than by
efficient angular momentum removal during collapse.  In this picture, disk
formation begins early, but the resulting disk size depends sensitively on
the initial angular momentum and its evolution during infall.   However,
measuring disk sizes in the youngest systems is observationally challenging,
and the inferred sizes may depend on the angular resolution, sensitivity,
optical depth, molecular or dust tracer used, and the adopted definition of
the disk boundary.

At the same time, the presence of systems such as HH~111, where significant
angular momentum loss is observed at smaller radii \citep{Lee2016HH111, Lin2025},
indicates that magnetic braking can become important under certain
conditions, possibly at later evolutionary stages.  Together, these results
suggest that while angular momentum is largely conserved during the early
phase of collapse, magnetic braking may play an increasing role as the
system evolves, leading to a diversity of disk properties.

\section{Conclusions}

I have investigated the evolution of specific angular momentum from core to
disk scales in three protostellar systems -- HH~212, HH~211, and B335 --
using observations and the framework of inside-out collapse.  Despite large
differences in magnitude, all three sources exhibit a similar radial
behavior: a power-law dependence at core scales transitioning to an
approximately constant value in the inner envelope.

This transition can be naturally understood within the framework of
inside-out collapse, in which
material in the collapsing region originates from a limited range of initial
radii and thus retains approximately constant specific angular momentum. 
The observed profiles are  broadly well described by the Terebey-Shu-Cassen (TSC)
model, with inferred collapse ages and infall rates yielding accreted masses
that are consistent with the central masses derived from kinematics,
allowing for a fraction of the material to be ejected by jets and winds.

For the three sources studied here,
the inferred midplane densities, accounting for the flattened structure of
magnetized cores, are also consistent with observations.  This supports a
picture in which magnetic fields primarily shape the geometry of the
collapse, producing flattened pseudodisk structures.
The observations are also consistent with 
approximate conservation of specific angular momentum during the early collapse phase.


Although the radial behavior is similar among the sources, the magnitude of
the specific angular momentum varies substantially.  In particular, the
small disk in B335 can be explained by its lower initial angular momentum,
although alternative explanations, such as stronger magnetic coupling or
different initial conditions, cannot be excluded, and observational
uncertainties in measuring disk sizes in the youngest systems should also
be taken into account.


Overall, these results suggest that inside-out collapse, with modest magnetic
modification, provides a plausible first-order description of angular
momentum evolution from core to disk scales, with angular momentum largely
conserved during the early stages of star formation
for the three sources
investigated in this work. A larger sample will be required to establish
the generality of this interpretation.

\begin{acknowledgements}


I thank the referee for the thoughtful and constructive comments, which
have substantially improved the manuscript. C.-F.  L.  acknowledges support
from the National Science and Technology Council of Taiwan under Grants No. 
112-2112-M-001-039-MY3 and 115-2112-M-001-058-MY3.

\end{acknowledgements}






\def\nat{Natur}

\end{document}